\documentclass{PoS}

\usepackage{amsmath}
\usepackage{braket}
\usepackage{subcaption}
\usepackage{multirow}
\usepackage{feynmf}
\usepackage{slashed}

\title{Beyond the Standard Model Kaon Mixing with Physical Masses.}

\ShortTitle{BSM kaon mixing at the physical point}
\author{Peter Boyle \\ 
	Higgs Centre for Theoretical Physics, University of Edinburgh, EH9 3FD, Edinburgh, UK \\
	E-mail: \email{paboyle@ph.ed.ac.uk}}
\author{Nicolas Garron\\
	Dept. of Mathematical Sciences, University of Liverpool, L69 3BX, Liverpool, UK \\
	E-mail: \email{Nicolas.Garron@liverpool.ac.uk}}
\author{Renwick James Hudspith \\ 
	York University, Department of Physics and Astronomy,
	Toronto, Ontario,
	Canada,
	M3J 1P3 \\
	E-mail: \email{renwick.james.hudspith@googlemail.com} 
}
\author{Andreas J{\"u}ttner \\
	School of Physics and Astronomy,
	University of Southampton,
	Highfield,
	Southampton, UK, SO17 1BJ \\
	E-mail: \email{juettner@soton.ac.uk}}
\author{\speaker{Julia Kettle}\\
        Higgs Centre for Theoretical Physics, University of Edinburgh, EH9 3FD, Edinburgh, UK\\
        E-mail: \email{J.R.Kettle-2@sms.ed.ac.uk}}
\author{Ava Khamseh \\
	Higgs Centre for Theoretical Physics, University of Edinburgh, EH9 3FD, Edinburgh, UK \\
	E-mail: \email{ava.khamseh@igmm.ed.ac.uk}}
\author{Justus Tobias Tsang \\
   Higgs Centre for Theoretical Physics, University of Edinburgh, EH9 3FD, Edinburgh, UK \\
	E-mail: \email{j.t.tsang@ph.ed.ac.uk}}

\abstract{We present results from a calculation of beyond the standard model (BSM) kaon mixing including data with physical light quark masses. 
We simulate $N_f=2+1$ QCD with Iwasaki gauge and domain wall fermion action on 8 ensembles, spanning 3 lattice spacings and pion masses from the physical value up to 430MeV.
The ratio of the BSM to standard model (SM) matrix elements are extracted from the correlation functions and renormalised using the RI-SMOM Rome-Southampton method with non-exceptional kinematics.
The results at the physical point continuum limit are found by performing a simultaneous continuum chiral extrapolation. 
In this work we gain consistency with our previous results and achieve a reduction in both the statistical and systematic error. }

\FullConference{The 36th Annual International Symposium on Lattice Field Theory - LATTICE2018\\
		22-28 July, 2018\\
		Michigan State University, East Lansing, Michigan, USA.}

\begin{document}

\section{Introduction}

        In the standard model kaon mixing has been recognised as an important area of study since the discovery of CP violation in the 
        $K_s$ regeneration experiment by Christenson et al. It is directly related to $\epsilon_k$ the degree of indirect CP violation within the standard model.

\begin{figure}[h]	
                \begin{center}
                        \begin{fmffile}{KaonMixing} 
                                        \begin{fmfgraph*}(80,35) 
                                                \fmfleft{i1,i2} 
                                                \fmfright{o1,o2} 
                                                \fmflabel{$\bar{d}$}{i2} 
                                                \fmflabel{$d$}{o1} 
                                                \fmflabel{$s$}{i1} 
                                                \fmflabel{$\bar{s}$}{o2} 
                                                \fmf{fermion}{i1,v1} 
                                                \fmf{fermion,tension=.5,label=$\bar{u},,\bar{c},,\bar{t}$,1.side=right}{v1,v3} 
                                                \fmf{fermion}{v3,o1} 
                                                \fmf{fermion}{o2,v4} 
                                                \fmf{fermion,tension=.5,label=$u,,c,,t$,1.side=right}{v4,v2} 
                                                \fmf{fermion}{v2,i2} 
                                                \fmf{photon,tension=0,label=$W$,1.side=left}{v1,v2} 
                                                \fmf{photon,tension=0,label=$W$,1.side=left}{v3,v4} 
                                                \fmfdotn{v}{4} 
                                        \end{fmfgraph*}     
                        \end{fmffile} 
                        \vspace{0.2cm}
                \caption{W exchange box diagram}
                \label{fig:kkmixing}
                \end{center}
        \end{figure}
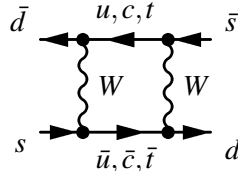

        Kaon mixing is a flavour changing neutral current (FCNC) process, occuring at one-loop. The process is dominated by box-diagrams, such as figure \ref{fig:kkmixing}, mediated by the neutral W boson.

        \vspace{0.5cm}

        \vspace{-0.3cm}
        The long-distance matrix element $\braket{\bar{K}^0|\hat{O}_1|K^0}$, where $\hat{O}_1$ is the effective four quark weak operator mediating the $\Delta S = 2$ transition in the SM, can be separated out from the short-distance wilson coefficients, by an operator product expansion (OPE).
        The calculation of the matrix element can be done with lattice QCD and has been the subject of many studies. It is now known to a high level of precision, as summarised in the FLAG report\cite{Aoki2017}. 
        Beyond the standard model, other matrix elements with differing Dirac structures are allowed. 
        We can construct a weak effective Hamiltonian of the possible operators:

\begin{equation}
        \mathcal{H}^{\Delta S = 2} = \sum_{i=1}^5 C_i(\mu) O_i(\mu) + \sum_{i=1}^3 \tilde{C_i}(\mu) \tilde{O_i}(\mu).
        \label{eq:ds2Hamilton}
    \end{equation}
    In our framework, where parity is conserved, the $\tilde O$ are redundant, therefore we only need to study the five operators $O_i$. In our framework we are interested only in the parity-even operators.
    In the so-called SUSY basis introduced in \cite{Gabbiani:1996hi}, the parity-even operators are,
    \begin{equation}
    \begin{split}
    O_1 &=(\bar{s}_a \gamma_{\mu}(1-\gamma_{5})d_a)(\bar{s}_b \gamma_{\mu}(1-\gamma_{5})d_b) \\
    O_2 &= (\bar{s}_a (1-\gamma_{5})d_a)(\bar{s}_b (1-\gamma_{5})d_b)\\
    O_3 &= (\bar{s}_a (1-\gamma_{5})d_b)(\bar{s}_b (1-\gamma_{5})d_a)\\
    O_4 &= (\bar{s}_a (1-\gamma_{5})d_a)(\bar{s}_b (1+\gamma_{5})d_b)\\
    O_5 &= (\bar{s}_a (1-\gamma_{5})d_b)(\bar{s}_b (1+\gamma_{5})d_a).\\
    \label{eq:opbasis_susy}
    \end{split}
    \end{equation}
There have been lattice studies of BSM kaon mixing by ETM\cite{Carrasco:2015pra}, SWME\cite{Jang:2015sla} and RBC-UKQCD\cite{Garron:2016mva}.


        We parametrise the BSM matrix elements as a ratio over the the SM matrix elements. 
        Following \cite{Babich:2006bh}, we define ratio parameters

        \begin{equation}
        R_i\bigg( \frac{m^2_P}{f^2_P} , a^2, \mu\bigg) = 
        \bigg[ \frac{f_K^2}{m_K^2} \bigg ]_{Exp.} 
        \bigg[ \frac{m_P^2}{f_P^2} 
        \frac{\braket{\bar{P}|O_i(\mu)|P} }{ \braket{\bar{P}|O_1(\mu)|P} } \bigg]_{Lat.}
        \end{equation}
        where $m_P$ and $f_P$ are the mass and decay constant of the pseudoscalar meson on the lattice. The term in ${m_P/f_P}$ is included to control the behaviour at the chiral limit, but
        at the physical point this parameterisation reduces to a direct ratio of the BSM to SM matrix element. 

\section{Simulation}

We use RBC-UKQCD's $N_f = 2+1$ gauge ensembles with the Iwasaki gauge action \cite{Iwasaki:1984cj,Iwasaki:1985we} and the domain wall fermion (DWF) action  of either M\"obius (M) \cite{Brower:2004xi,Brower:2005qw,Brower:2012vk} or Shamir (S) \cite{Shamir:1993zy} kernel.
In our previous study of BSM kaon mixing \cite{Garron:2016mva} there were only 2 lattice spacings and no data with physical pion masses.
This work improves upon our previous work by including a third lattice spacing \cite{Boyle:2017jwu}. 
The pion masses span from the physical, on two new ensembles, up to 430 MeV. We use Z2 Gaussian wall (Z2GW) sources with smearing for all ensembles save C1(2) which are Z2 wall (Z2W) sources.  A summary of simulation parameters is given in table \ref{tab:enspar}.

\begin{table}[h]
        \caption{Summary of the ensembles used in this work. 
		      C, M and F stand for coarse, medium and fine, respectively. M and S stand for M\"obius and Shamir kernels, respectively. }
\label{tab:enspar}
\resizebox{\textwidth}{!}{
        \begin{tabular}{c | c c | c c c c c | c c c c}
                \hline
                \hline
                name & $L/a$ & $T/a$ & kernel & source & $a^{-1}[\textrm{GeV}]$ & $m_\pi[\textrm{MeV}]$ & $n_{configs}$ & $am_l^{uni} $  & $am_s^{sea}$ & $am_s^{val}$ & $am_s^{phys}$  \\
                \hline
                C0 & $48$ & $96$ & M & Z2GW & 1.7295(38) & 139 & 90 & 0.00078 & 0.0362 & 0.0358 &  0.03580(16) \\
                C1 & $24$ & $64$ & S & Z2W & 1.7848(50) & 340 & 100 & 0.005 & 0.04 & 0.03224 & 0.03224(18) \\
                C2 & $24$ & $64$ & S & Z2W & 1.7848(50) & 430 & 101 & 0.01 & 0.04 & 0.03224 & 0.03224(18) \\
                \hline
                M0 & $64$ & $128$ & M & Z2GW & 2.3586(70) & 139 & 82 & 0.000678& 0.02661 & 0.0254 &0.02539(17) \\
                M1 & $32$ & $64$ & S & Z2GW & 2.3833(86) & 303 & 83 & 0.004 & 0.03 & 0.02477 & 0.02477(18) \\
                M2 & $32$ & $64$ & S & Z2GW & 2.3833(86) & 360 & 76 & 0.006 & 0.03 & 0.02477 & 0.02477(18) \\
                M3 & $32$ & $64$ & S & Z2GW & 2.3833(86) & 410 & 81 & 0.008 & 0.03 & 0.02477 & 0.02477(18) \\
                \hline
                F1 & $48$ & $96$ & M & Z2GW & 2.774(10) & 234 & 98 & 0.002144 & 0.02144 & 0.02132 & 0.02132(17)  \\
                \hline
                \hline
        \end{tabular}
}
\end{table}

\begin{figure}
	\centering
 \begin{subfigure}[t]{0.4\textwidth}
        \includegraphics[width=\textwidth]{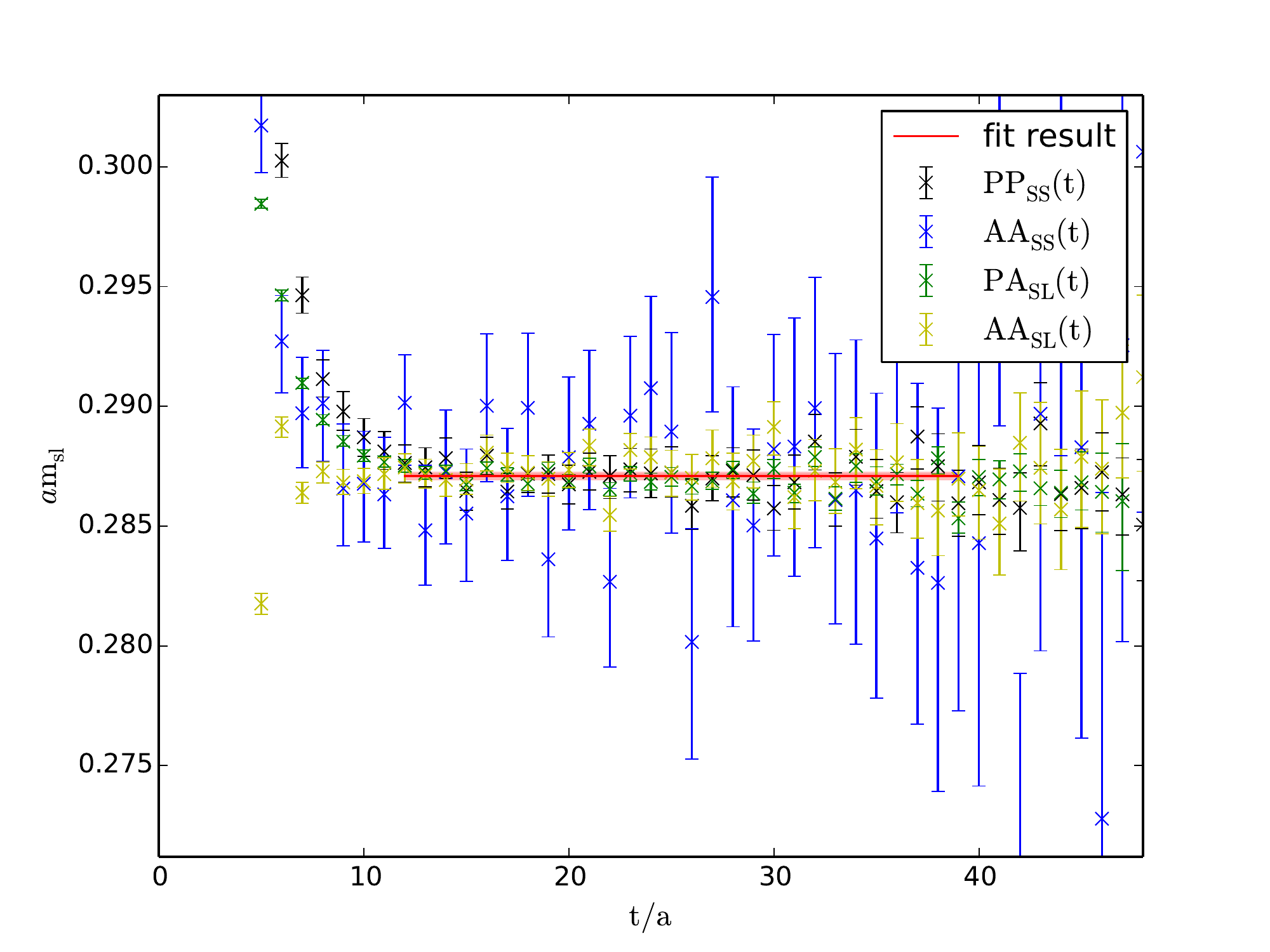}
 \end{subfigure}
\hspace{0.5cm}
 \begin{subfigure}[t]{0.4\textwidth}
        \includegraphics[width=\textwidth]{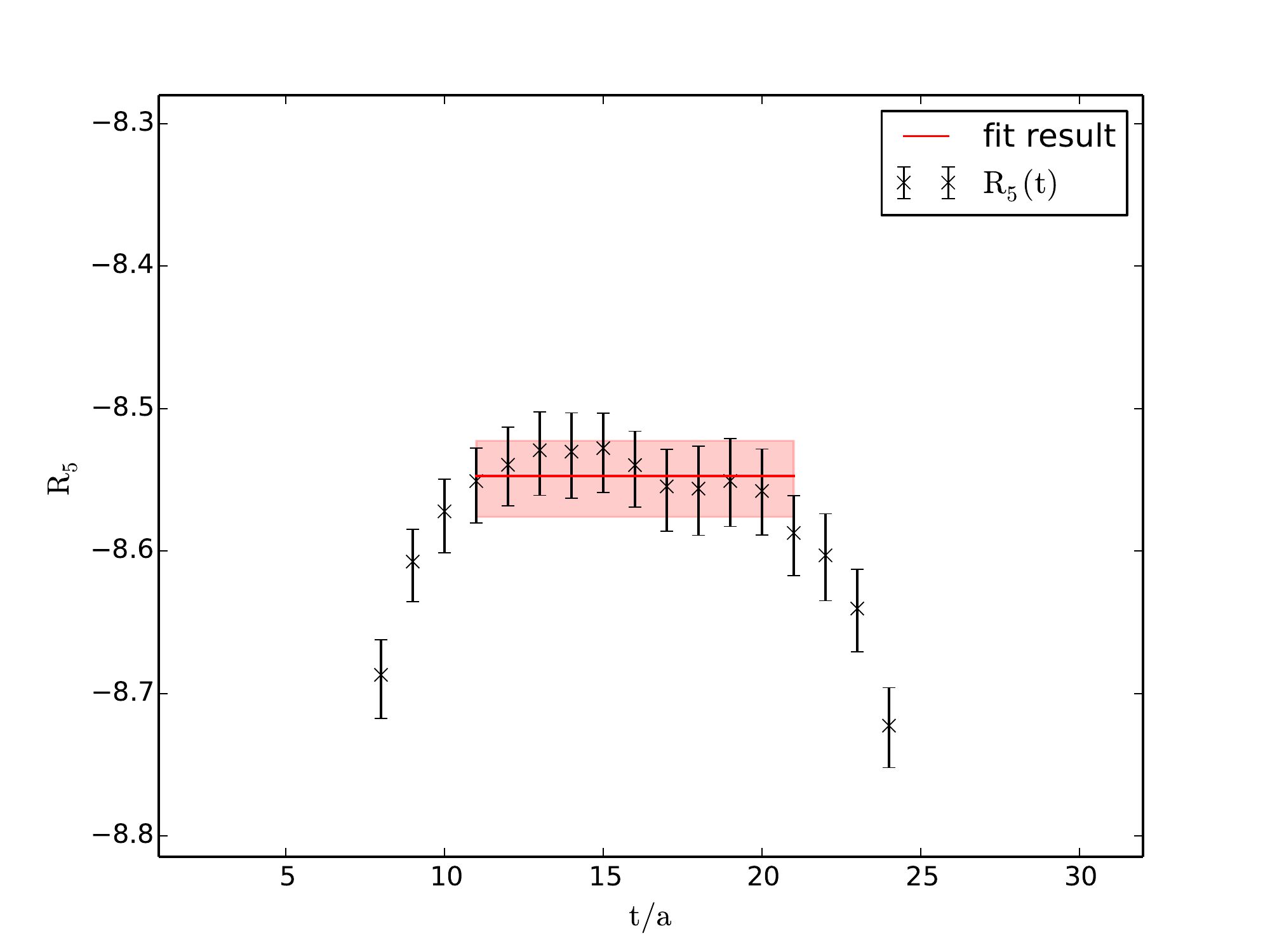}
 \end{subfigure}
 \caption{Examples of the correlator fits. On the left is a simultaneous fit of multiple channels of the two-point correlation function (for ensemble C0) to the ground-state to extract the effective mass. On the right is a fit of the ratio of a BSM to SM three-point correlation function for M0. }
 \label{fig:corrFit}
\end{figure}

We extract the ratio parameters and pseudoscalar masses and decay constants with a ground state fit of the three-point and two-point correlation functions respectively.
For the two-point correlation functions we simultaneously fit combinations of pseudoscalar and axial current channels.
We show some examples of the correlation functions fits in figure \ref{fig:corrFit}.

\vspace{-0.2cm}

\section{Renormalisation}

The renormalisation is performed using the non-perturbative Rome-Southampton method \cite{Martinelli:1994ty} with non-exceptional kinematics \cite{Sturm:2009kb} (RI-SMOM). 
We can then convert our results to the more useful  $\overline{\textrm{MS}}$ scheme using the one-loop perturbative matching as given in \cite{Boyle:2017skn}.
\begin{equation}
O_i (\mu)^{\overline{\textrm{MS}}} = R^{\overline{\textrm{MS}} \leftarrow RI-SMOM}(\mu) \lim_{a \rightarrow 0 } [Z_{ij}^{RI-SMOM}(\mu,a)\braket{O_j(a)}].
\end{equation}
The scale is defined by the momenta and should fall within the Rome-Southampton window; \hspace{0.1cm} $\Lambda^2_{QCD} \ll \mu^2 \ll (\pi/a)^2$.
The upper limit is given by the lattice cut-off and the lower limit ensures accurate perturbative matching to $\overline{\textrm{MS}}$.
We use non-exceptional momenta ($p_1^2 = p_2^2 = (p_1-p_2)^2$ and $p_1 \neq p_2$) so that the infrared effects are more suppressed \cite{Aoki:2007xm}.
In this work, two schemes, $(\gamma_\mu,\gamma_\mu)$ and $(\slashed{q},\slashed{q})$, are used, they are distinguished by their projectors.  Definitions and further details can be found in \cite{Boyle:2017skn}. 
The renormalisation is calculated on a small subset of configurations and results for C1/2 and M1/2/3 are given in \cite{Boyle:2017skn}. The renormalisation factors for C0, M0 and F1 will be published in a forthcoming paper. 

We perform the renormalisation at 2GeV and 3GeV, both of which fall within the Rome-Southampton window.
The higher scale however is more susceptible to discretisation effects, particularly on the coarser lattices, while the lower scale will have a larger error in the perturbative matching.
We can scale an operator renormalised at one scale to another using a scale evolution matrix:
\begin{equation}
\sigma(\mu_1,\mu_2) = \lim_{a^2 \rightarrow 0 } Z(\mu_1,a) Z^{-1}(\mu_2,a).
\end{equation}
In this way we renormalise at 2GeV where the discretisation effects are small, and use the scale-evolution matrix, calculated on only the medium and fine ensembles, to scale our results to 3GeV where the perturbative matching has a smaller error. 

\section{Extrapolation}

We perform a simultaneous chiral continuum extrapolation according to the global fit form,
\begin{equation}
        Y\bigg( a^2,\frac{m_{ll}^2}{f_{ll}^2} \bigg) = Y\bigg( 0,\frac{m_{\pi}^2}{f_{\pi}^2}, 0 \bigg) \bigg[ 1 + \alpha_i a^2 + \beta_i \frac{m_{ll}^2}{f_{ll    }^2} \bigg],
        \label{eq:fitform}
\end{equation}
linear in $a^2$ and in $m_{ll}^2/f_{ll}^2$. 
The lattice spacings were determined in a global fit \cite{Blum:2014tka} (which was updated in \cite{Boyle:2017jwu} to include F1) including some of the same ensembles as in this work. Hence there is a small correlation. However, the error on the lattice spacings is of order 0.5\% and the largest discretisation effect seen in any of the data points is of order 15\% or less, leading to an effect of order 0.01\%. Thus we neglect this effect. 
We perform the fit using a $\chi^2$ minimisation, where the $\chi^2$ is calculated using only the error in y-axis. 
The gradient of the fit (scaled by the error in the mass or lattice spacing) is small compared to the ratio parameter error in both cases. Therefore the difference in our values of $\chi^2$ from those we would obtain including the errors in the pion mass and lattice spacing is negligible.

\section{Results}

\begin{figure}[h!]
	\centering
        \begin{subfigure}{0.44\textwidth}
                \includegraphics[width=\textwidth]{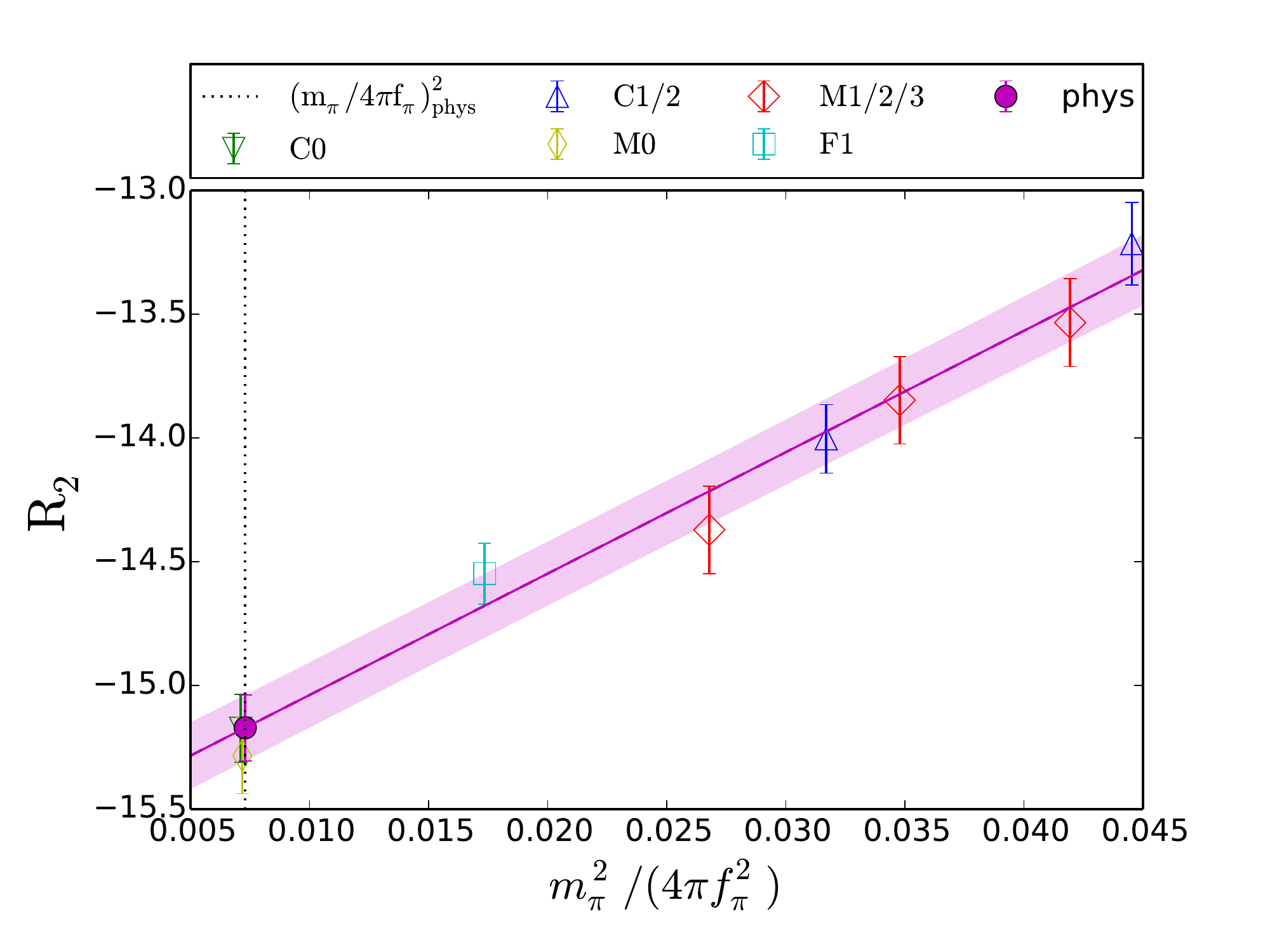}
        \end{subfigure}
        \begin{subfigure}{0.44\textwidth}
                \includegraphics[width=\textwidth]{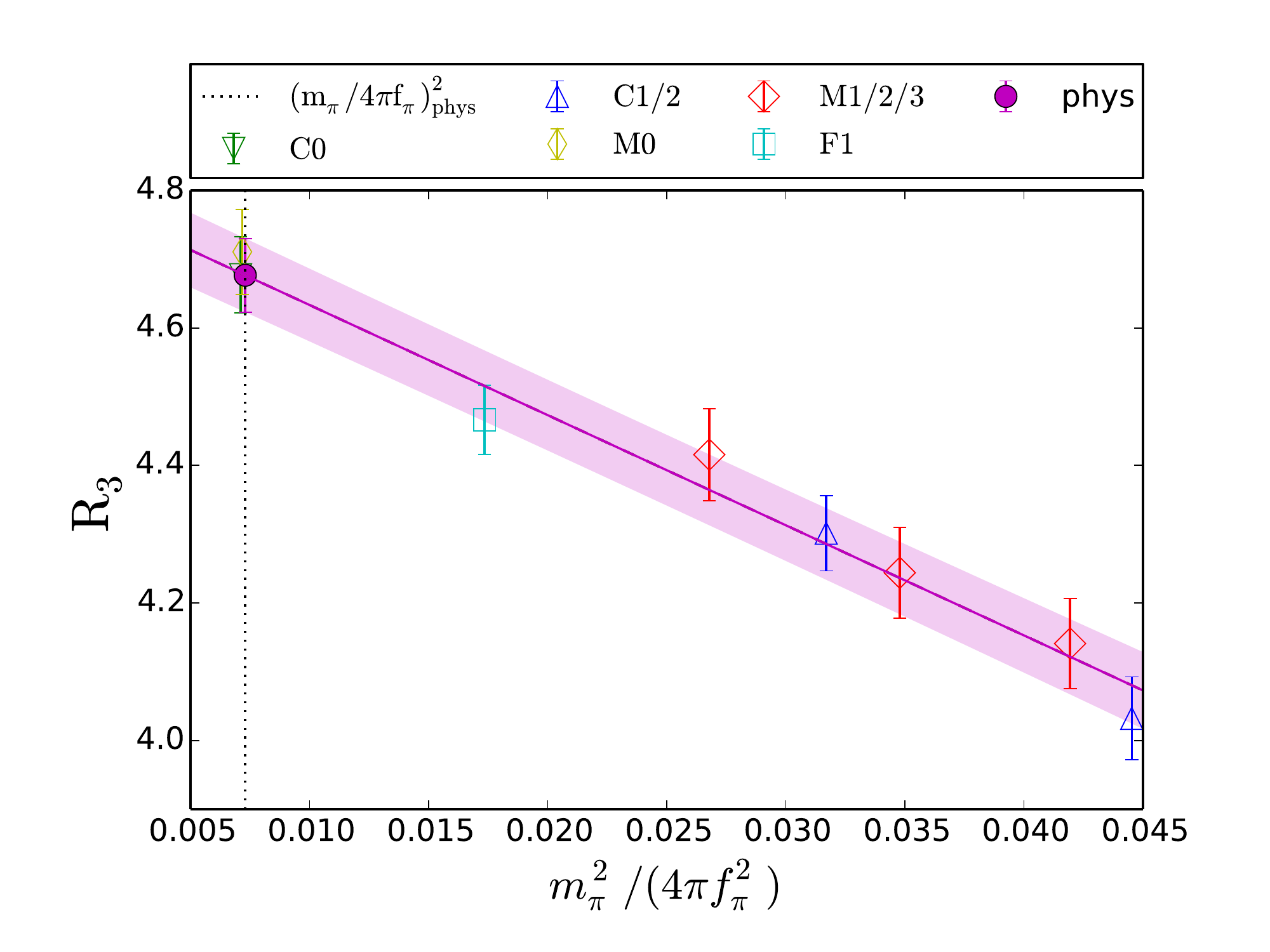}
        \end{subfigure}

        \begin{subfigure}{0.44\textwidth}
                \includegraphics[width=\textwidth]{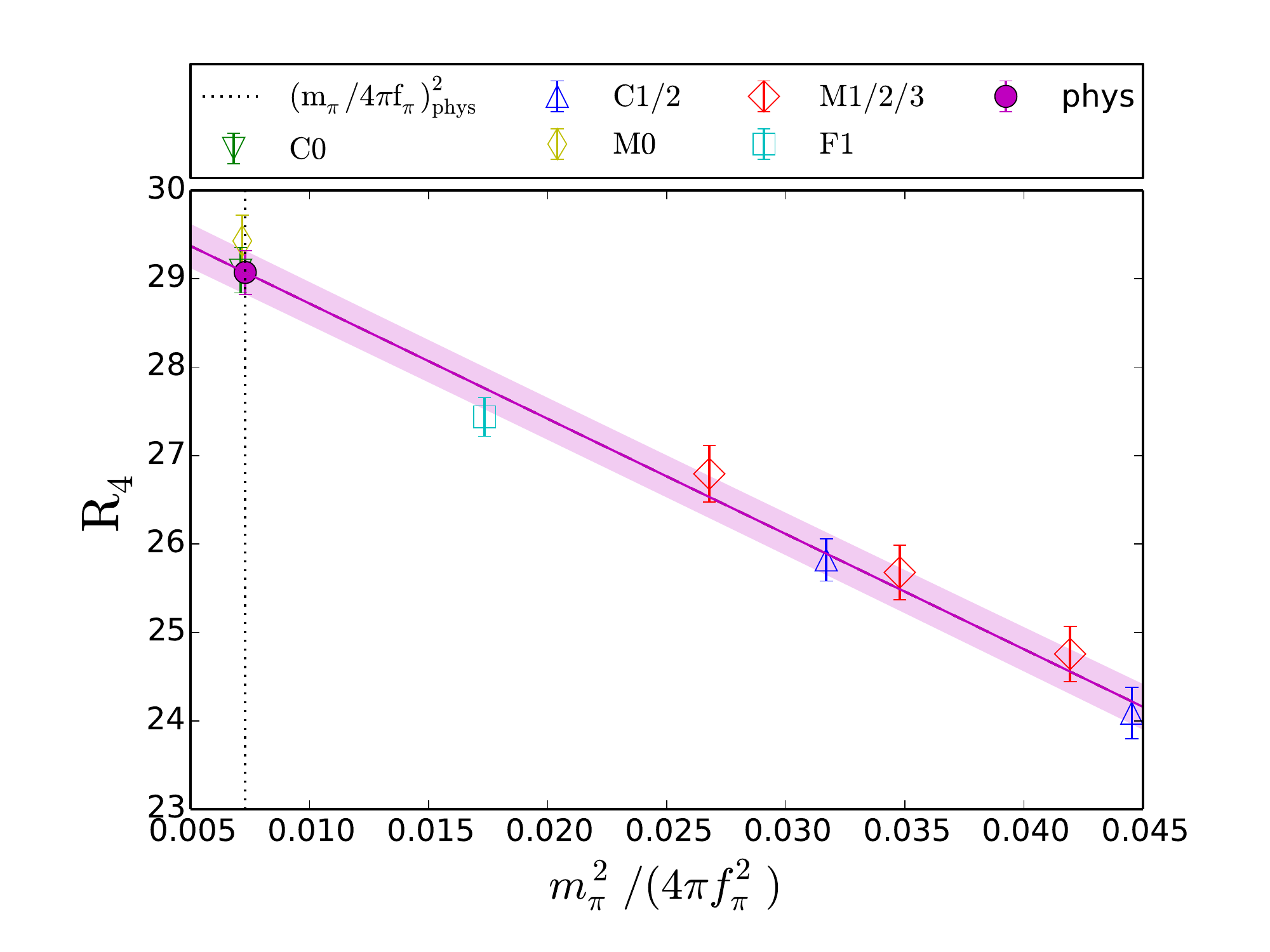}
        \end{subfigure}
        \begin{subfigure}{0.44\textwidth}
                \includegraphics[width=\textwidth]{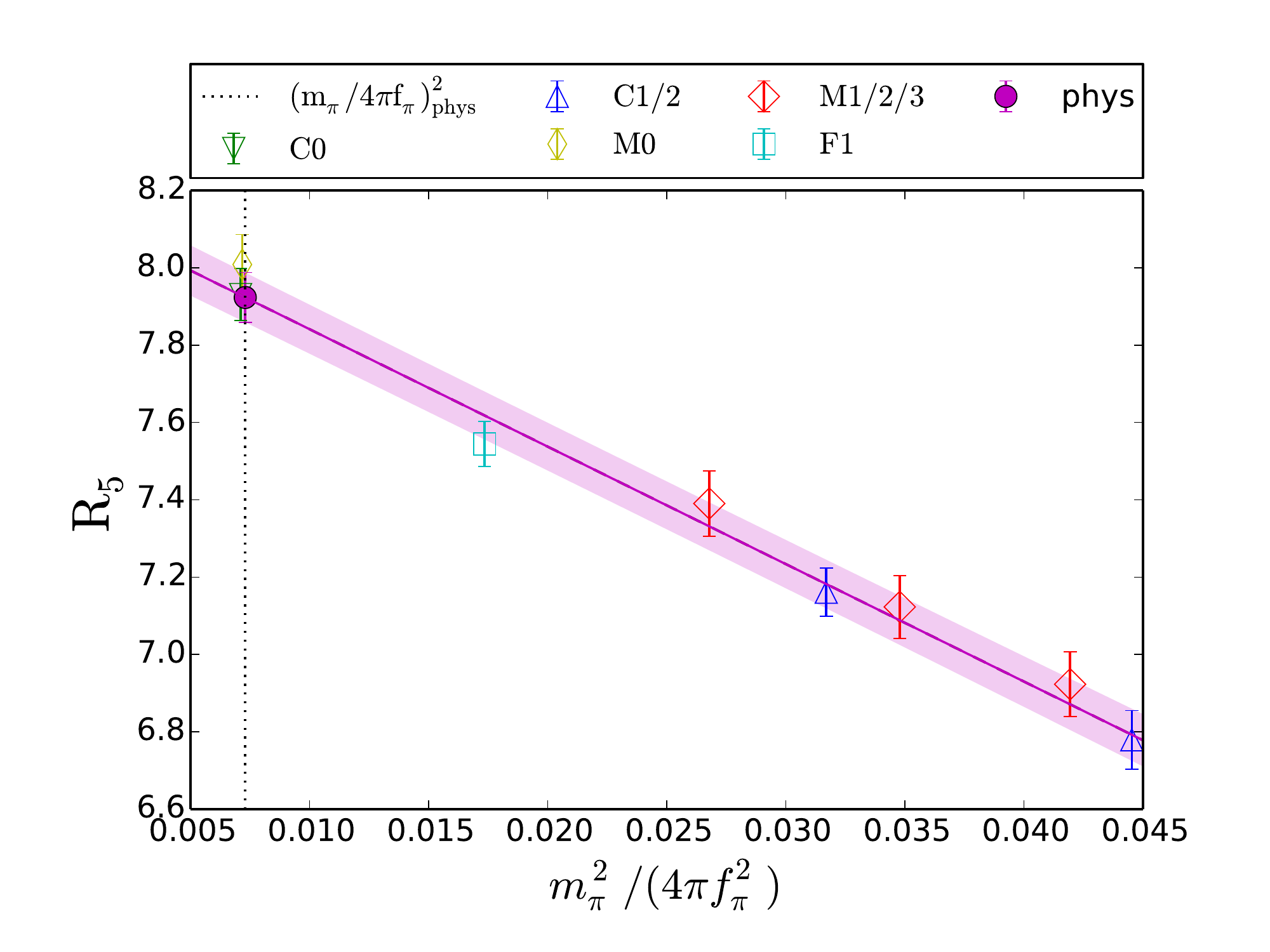}
        \end{subfigure}

        \caption{The chiral extrapolation of the ratio parameters, renormalised in SMOM$^{(\gamma,\gamma)}$ at 2GeV, is shown. The results were gained from a combined 
        chiral continuum fit and the data points have been corrected to the continuum using the parameters gained from the fit.\vspace{-0.15cm} }
        \label{fig:results}
\end{figure}

In figure \ref{fig:results} we show the fits of the ratio parameters at 2GeV in the intermediate renormalisation scheme SMOM$^{(\gamma_\mu,\gamma_\mu)}$.

We present our final results in $\overline{\textrm{MS}}$ at 3GeV after being renormalised in RI-SMOM$^{(\gamma,\gamma)}$ at 2GeV, step-scaled \cite{Arthur:2010ht} to 3GeV and matched to $\overline{\textrm{MS}}$.
We present a full error budget in table \ref{tab:error}. The statistical error is small typically of order 1\%.
Our fit ansatz does not follow chiral perturbation theory, and is instead linear in $m_{ll}^2/f_{ll}^2$. We also perform fits according to $\chi{PT}$, in which the fit form differs only in the inclusion
of a logarithmic term. By comparing the difference between the results from the linear and chiral ansatz we are able to estimate
the error arising from the the global fitting of the mass dependence. This is labelled as "Chiral." in table \ref{tab:error}.

Given the inclusion of the third lattice spacing the continuum extrapolation is well controlled. This inclusion of the third lattice spacing, of course necessitates understanding the mass dependence in a global fit. We have had no prior evidence of higher order lattice artefacts in our light hadron physics, suggesting that hadronic quantities display modest scaling violations. However we expect there may be discretisation errors arising from the hard off shell renormalisation, since the scale of these is set by a momentum deliberately chosen to lie in the Rome Southampton perturbative window. with no observation of higher order lattice artefacts. By choosing to renormalise at 2GeV and step-scale to 3GeV using a scale-evolution matrix determined excluding the coarse lattice, we expect to have minimised the effect of discretisation errors. 
We compare the step-scaled results used for our central values and the results renormalised directly at 3GeV to provide an estimate of the magnitude of the discretisation effects. 
These are labelled in table \ref{tab:error} as "Discr.".

We estimate that the dominant source of systematic error in our results is in fact the perturbative matching at one loop to MS arising from the truncation in the series. 
We expect, in the continuum limit, the final result to be independent from the intermediate renormalisation scheme if matching is performed to all orders.
Since, we have results in two renormalisation schemes (γμ , γμ ) and (q/,q/), which should differ only in the truncation of the perturbative series in the matching, we can
use these to assess perturbative error in the extrapolated continuum value after our global fit. We take the difference between the results as an estimate of the magnitude 
of the perturbative matching error. This is labelled as "P.T.". This is the largest source of error, but for the ratios it is still of order 3

\begin{table}[h!]
        \centering
        \caption{Central values and error budget for our final results renormalised at $\mu=2\textrm{GeV}$, step-scaled to, and converted to $\overline{\textrm{MS}}$ at, 3GeV.  For our results converted to 
        $\overline{\textrm{MS}}$ our central value is obtained using SMOM$^{(\gamma_\mu,\gamma_\mu)}$ as the intermediate scheme.}
        \label{tab:error}
        \begin{tabular}{c | c | c c c c }
            \hline \hline
        scheme &  & $R_2$ & $R_3$ & $R_4$  & $R_5$ \\
        \hline
        \multirow{7}{*}{$\overline{\textrm{MS}} \leftarrow \textrm{SMOM}^{(\gamma_\mu,\gamma_\mu)}$} 
        & central 
        & -18.83 & 5.815 & 41.58 & 10.814 \\
        \cline{2-6}
        & Stat.  
        & 0.90\% & 1.08\% & 0.89\% & 0.82\% \\
        & Chiral. 
        & 0.90\% & 0.94\% & 1.65\% & 1.61\% \\
        & Discr  
        & 0.85\% & 0.67\% & 1.55\% & 1.99\% \\
        &  P.T. 
        & 2.78\% & 1.32\% & 2.47\% & 2.89\% \\
        \cline{2-6}
        & Total.  
        & 3.04\% & 1.75\% & 3.35\% & 3.86\% \\
        \hline \hline
        \end{tabular}
\end{table}
\vspace{-0.2cm}
\section{Conclusions}

In this work we have improved upon our previous calculation of the BSM kaon mixing ratio parameters by including a third lattice spacing, and data directly at the physical point.
The results we have gained are consistent with our previous work, as shown in table \ref{tab:compmyresults}, and have two-fold reduced statistical errors. 
We have calculated the full error budget including systematic errors. For all the ratio parameters these are either of the same order as in the previous work or have been reduced, as we have a better controlled continuum and chiral extrapolation given the inclusion of fine lattice and physical pion mass data. 
The precision gained here is sufficiently high that to make meaningful gains any further work on this topic should consider isospin breaking effects
and calculate the two-loop matching coefficients (a computation is already underway, see talk by Kvedaraite at this conference).  Earlier stages of this were presented in previous Lattice conferences \cite{Boyle:2017ssm}. 
These results will be published in full, alongside the bag parameters and the matrix elements, in a forthcoming paper. 

\begin{table}
        \centering
	\caption{Comparison of the results of this work in $\overline{\textrm{MS}}(\mu=3\textrm{GeV})$ alongside our collaboration's
	previous results presented in \cite{Garron:2016mva}.}
	\label{tab:compmyresults}
	\begin{tabular}{ c | c  c }
		\hline \hline
		&  \multicolumn{1}{c}{RBC-UKQCD16\cite{Garron:2016mva}} & This Work \\
		\hline \hline
		$N_f$ = & 2+1  & 2+1 \\
		scheme & RI-SMOM  & RI-SMOM\\
		\hline
		\hline
		$R_2$  & -19.48(44)(52)  &   -18.83(17)(55) \\
		$R_3$  &  6.08(15)(23)   &   5.815(63)(125)  \\
		$R_4$  &  43.11(89)(230) &   41.58(37)(119) \\
		$R_5$  &  10.99(20)(88)  &   10.81(9)(37) \\
		\hline \hline
	\end{tabular} 
\end{table}

\section{Acknowledgements}
This work used the DiRAC Blue Gene Q Shared Petaflop system and the Extreme Scaling service operated by the Edinburgh Parallel Computing Centre on behalf of the STFC DiRAC HPC Facility (www.dirac.ac.uk).
The former equipment was funded by BIS National E-infrastructure capital grant ST/K000411/1, STFC capital grant ST/H008845/1, and STFC DiRAC Operations grants ST/K005804/1 and ST/K005790/1, and the latter by BEIS capital funding via STFC capital grant ST/R00238X/1 and STFC DiRAC Operations grant ST/R001006/1. DiRAC is part of the National e-Infrastructure. This research has received funding from the SUPA student prize scheme, Edinburgh Global Research Scholarship, and an STFC studentship.

\bibliographystyle{ieeetr}
\bibliography{lattice2018}

\end{document}